# Study of Twistronics Induced Superconductivity in Twisted Bilayer Graphene

Rajendra Paudel*, Nabin Upadhya Dhakal and Nurapati Pantha

Central Department of Physics, Tribhuvan University, Kirtipur 44618, Kathmandu, Nepal

*Corresponding author: rajendra.795511@cdp.tu.edu.np

## Abstract:

This work investigates the electronic properties of twisted bilayer graphene (TBG) through computational calculations, with the aim of understanding the emergence of flat bands and conditions favorable for superconductivity close to the magic angle. This study utilizes $\vec{k} \cdot \vec{p}$ continuum model and the low-energy Hamiltonians were derived from angle-dependent datasets provided by Carr et al. Using this model, band structure, density of states (DoS), and Fermi velocity were systematically calculated across a range of twist angles. The calculations were performed by discretizing high-symmetry paths in the moiré Brillouin zone for band structure calculations, uniformly sampled using square grid size for DoS analysis, and employing finite difference methods to evaluate Fermi velocity near the Dirac points. The results identify a narrow magic angle window around $\theta \approx 0.98°–1.00°$, where bands become nearly dispersionless, the DoS exhibits a sharp peak, and the Fermi velocity is strongly suppressed. This computational framework does not directly predict superconductivity but rather establish the electronic foundation for exploring the flat-band physics and correlation-driven phenomena like unconventional superconductivity in Twisted Bilayer Graphene.

## 1. Introduction

Twisted bilayer graphene (TBG) has emerged as a platform for exploring strongly correlated electronic phases, following the groundbreaking discovery of superconductivity near the "magic angle" of approximately 1.1◦ [1, 2]. When two graphene layers are rotated relative to each other by a small angle, the resulting moiré pattern induces a reconstruction of the band structure, leading to the formation of nearly flat bands near the Fermi level [3]. These flat bands dramatically enhance the density of states (DoS) and promote strong electron-electron interactions, giving rise to unconventional superconductivity, Mott-like insulating states, and topological phases [4]. The field of twistronics explores how electronic properties in V0an der Waals materials can be tuned via interlayer twist angle [1]. Continuum $\vec{k} \cdot \vec{p}$ models, derived from first-principles calculations, have proven particularly effective in capturing the low-energy electronic behavior of TBG, especially when lattice relaxation and interlayer coupling are included [5, 6].

The concept of twistronics builds on the idea that twisting two graphene layers creates a long-wavelength moiré pattern that dramatically modifies the band structure [3, 2]. In particular,

Bistritzer and MacDonald (2011) showed that at special "magic" twist angles the Dirac velocity vanishes and the lowest moiré band becomes nearly flat [3]. These flat bands have an enormously enhanced density of states (DoS), setting the stage for strong electron–electron interactions and correlated phases.

In 2018, Cao et al. reported that at the first magic angle (1.1) TBG exhibits correlated insulating states and superconductivity [1,2]. In devices tuned by gating, partial filling of flat band produced an insulating phase with Mott-like character. Upon electrostatic doping, zero-resistance superconducting domes appeared with $T_c$ upto $\sim 1.7K$. The superconducting phase diagram resembled that of cuprates. These landmark experiments demonstrated that a carbon-based 2D material can realize strong-correlation physics in the absence of magnetic fields [1,2]. Around the same time, theorists such as Po et al. developed effective Hubbard models on the moiré lattice andproposed a valley-coherent Mott insulator at half-filling [7].

In 2019, Tarnopolsky et al. introduced a simplified continuum model that yielded perfectly flat bands at magic angles [8], and Yankowitz et al. showed that superconductivity could be tuned by pressure [9]. Lu et al. fabricated ultra-uniform devices and discovered multiple superconducting domes, including those near neutrality and ±1 fillings, demonstrating the ubiquity of correlation effects across fillings [10].

By 2020, Stepanov et al. showed that superconductivity can exist without a strong correlated insulator by weakening Coulomb interactions [11]. Scanning tunneling microscopy (STM) work by Wong et al. observed "cascades" of band transitions at each integer filling [12], confirming interaction-driven sub-band formation. Meanwhile, first-principles studies emphasized the role of lattice relaxation in producing correct flat-band physics [13], and Koshino and Nam highlighted the strong electronphonon coupling near the magic angle [14].

In 2021, nano-ARPES studies by Lisi et al. and Utama et al. directly imaged the flat minibands near the magic angle [15, 16]. Mesple et al. emphasized the dominant role of heterostrain in modulating the flat-band condition [17]. In 2024, Li et al. tracked the band evolution from 1.1 to 2.6, confirming that relaxation increases sharply near the magic angle and directly affects the band dispersion [18].

Most recently, Yu et al. (2023) argued that hybridization between valence and conduction bands underlies the magic-angle band flattening [19]; these developments, along with growing capabilities in twist-angle control and spectroscopy, establish TBG as a model platform for exploring correlated phases, unconventional superconductivity, and flat-band physics in two dimensions.

In this work, we use the continuum-model approach and perform a high-resolution angular sweep of TBG from 0.80° to 1.35°, with fine steps around the critical regime. Our analysis reveals that the magic angle is not a single value but a narrow window between 0.98°–1.00°, within which three key observables converge: band flattening, sharp DoS enhancement, and Fermi velocity suppression. We further carry out a comparative study of relaxed and non-relaxed structures, demonstrating the essential role of atomic relaxation in deepening the flat-band condition. Finally, by documenting our Python-C++ computational workflow based on kp_tblg, we provide an accessible and replicable framework for researchers to probe moiré-induced superconductivity in TBG.

## 2. Method of Analysis

This study employs a computational approach based on the $\vec{k} \cdot \vec{p}$ continuum model for twisted bilayer graphene (TBG), utilizing the dataset and implementation by Carr et al. [20]. The primary objectives are to investigate the emergence of flat bands, calculate DoS, and determine Fermi velocities. All simulations were carried out on a Linux-based environment using Python 3.10 with scientific libraries such as NumPy and Matplotlib, along with CMake and g++ for compiling the C++ backend of the kp_tblg code. The modular structure of the code allows other researchers to replicate the findings, provided they have access to the same computational tools and dataset.

### 2.1 Model Framework

We use low-energy Hamiltonian of TBG near the Dirac points, based on the $\vec{k} \cdot \vec{p}$ continuum model originally proposed by Bistritzer and MacDonald and later extended by Carr *et al.* The total Hamiltonian incorporates both intralayer Dirac contributions and interlayer tunneling terms modulated by the moiré pattern [5], expressed as

$$H(\vec{k}) = \begin{pmatrix} h_1(\vec{k}) & T(\vec{r}) \\ T^\dagger(\vec{r}) & h_2(\vec{k}) \end{pmatrix} \quad (1)$$

where, $h_{1,2}$ represents Dirac Hamiltonians for top and bottom graphene layers, $T(\vec{r})$ captures the spatially varying interlayer coupling. Simulations were performed using the open-source kp_tblg code [20], which implements this model in both C++ and Python using the pybind11 interface.

### 2.2 Band Structure Calculation

The band structure was computed along the high-symmetry path of the moiré Brillouin zone (MBZ) defined by the sequence $K_m \rightarrow \Gamma_m \rightarrow M_m \rightarrow K_m$. These high-symmetry points ($\Gamma_m$: center, $K_m$: corner, $M_m$: edge midpoint) were obtained using the getGamma(), getK(), and getM() functions in the kp_tblg package [20]. Each segment of the path was discretized with 40 k-points, which we verified to be sufficient to produce smooth band dispersions. Increasing to 80 k-points did not noticeably change the results. At every k-point, the continuum Hamiltonian was constructed and diagonalized, with both k and –k included to maintain time-reversal symmetry. Eigenvalues were then shifted relative to the Fermi level, defined as the median eigenvalue at $\Gamma_m$. The resulting band structures were plotted with clear labeling of MBZ symmetry points to distinguish them from the monolayer graphene Γ, K, and M.

## 2.3 Density of States (DoS)

The DoS was computed by uniform sampling of the moiré Brillouin zone with a square grid. In the calculations, a 100×100 grid (10,000 k-points) was employed. At each k-point, the continuum Hamiltonian was diagonalized to obtain the eigenvalue spectrum, which was then shifted relative to the Fermi energy defined at the $\Gamma_m$ point.

$$\text{DoS}(E) \propto \frac{1}{N_k} \sum_{i=1}^{N_k} \delta(E - E_i) \tag{2}$$

The eigenvalues were accumulated into a histogram with 400 bins across the range –0.2 eV to +0.2 eV, corresponding to an energy resolution of 0.5 meV per bin. To reduce numerical noise, Gaussian broadening with σ = 1 meV was applied. The DoS was normalized to physical units of $eV^{-1} \cdot nm^{-2}$ by dividing the histogram counts by the bin width, the number of sampled k-points, and the moiré supercell area:

$$A_m = \frac{\sqrt{3}}{2} L_m^2 \tag{3}$$

where, $L_m$ is moiré periodicity.

This procedure was repeated for several twist angles to examine the evolution of the DoS profile, with particular focus on the emergence of a pronounced peak near the magic angle, indicative of flat band formation.

To ensure numerical accuracy, a convergence test of the density of states was carried out with k-point grids up to 120×120. The DoS peak position was found to converge within 1 meV, while the peak height varied by less than 3% beyond a 100×100 grid. Therefore, a 100×100 k-grid was adopted for all subsequent calculations as an optimal balance between accuracy and computational cost.

## 2.4 Fermi Velocity Estimation

The Fermi velocity ($v_F$) was obtained from the slope of the low-energy band dispersion in the vicinity of the moiré Dirac point. For each twist angle, the position of the moiré ($K_m$) point was extracted using the kp.getK() routine of the kp_tblg package. At this point, the continuum Hamiltonian was diagonalized to yield the eigenvalue spectrum. To estimate the group velocity, the band energy was evaluated at two nearby $k$-points displaced by a small step ($\delta k$) along the $k_x$ −direction. The velocity was then computed as:

$$v_F = \frac{1}{\hbar} \frac{E_1 - E_0}{\delta k} \tag{4}$$

where, $E_0 = E(K_m)$, $E_1 = E(K_m + \delta k\, \hat{x})$, $\delta k = 1 \times 10^{-4} (\text{Å})^{-1}$ and $\hbar = 6.582 \times 10^{-16} eV.s$

Finally, the computed Fermi velocity was normalized by the monolayer graphene reference value $v_0 = 1 \times 10^6$ m/s and the results are reported in the form $v_F/v_0$.
This procedure was repeated over a dense set of twist angles between 0.42° and 5.08°, using finer angular resolution near the magic angle. The Fermi velocity is compared for both relaxed and non-relaxed structures using separate datasets.

## 3. Theoretical Background

### 3.1 Bilayer and Twisted Bilayer Graphene

Graphene monolayers are stacked in two layers to form bilayer graphene (BLG). The Bernal (AB) stacking configuration, in which a top layer sublattice is situated precisely above a bottom layer sublattice, is the most stable. This alters monolayer graphene's electrical characteristics by breaking its sublattice symmetry. Unlike monolayer graphene's linear dispersion, AB-stacked BLG exhibits parabolic low-energy bands near the Dirac point, and its charge carriers behave as massive chiral quasiparticles. The effective low-energy Hamiltonian is [21]:

$$H_{eff} = \frac{-1}{2m^*} \begin{pmatrix} 0 & (\pi^\dagger)^2 \\ \pi^2 & 0 \end{pmatrix} \tag{5}$$

where, $\pi = \hbar(k_x + ik_y)$, $\pi^\dagger = \hbar(k_x - ik_y)$ *and* $m^*$ is the effective mass determined by interlayer coupling $\gamma_1$.

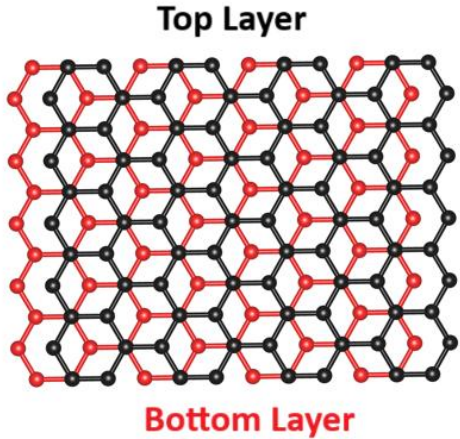

**Figure 3.1.1:** Illustration of AB stacking of graphene, the bottom layer is red, whereas the top layer is in black color which is created by using VESTA.

In contrast, AA stacking retains the Dirac cones of individual layers, leading to a metallic band structure due to shifted cones [22]. Symmetry plays a critical role in BLG. For example, a perpendicular electric field breaks inversion symmetry in AB-stacked BLG, opening a tunable band gap [23]. This tunability becomes especially relevant in TBG, where the layers are rotated by a small angle $\theta$, forming a long-wavelength Moiré superlattice [24]. The moiré periodicity is given by:

$$L_M = \frac{a}{2sin(\theta/2)} \tag{6}$$

where, $a = 2.46 A°$ as the graphene lattice constant. For small $\theta \lesssim 2°$, $L_M$ spans several nanometers, drastically altering the electronic structure.

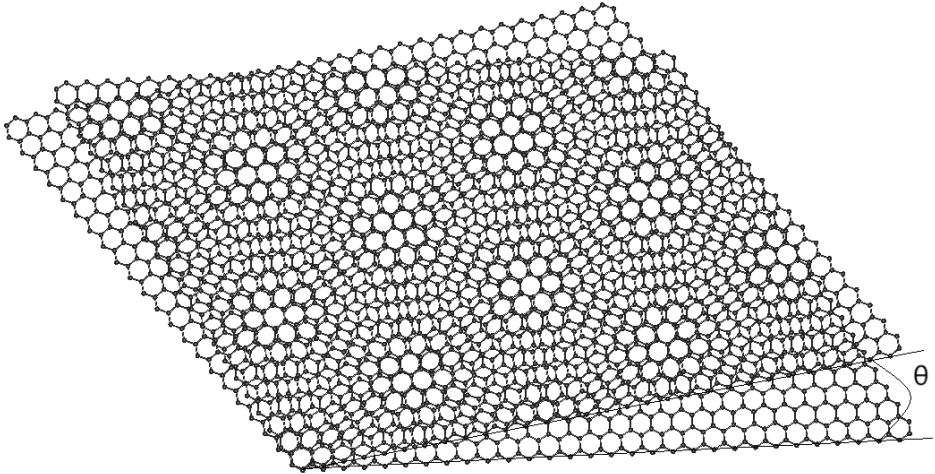

**Figure 3.1.2:** Moiré pattern is formed in TBG with twist angle $\theta$ which contain AA, AB and SP stacking, created using VESTA.

Within this superlattice, local stacking variations such as AA, AB, and SP regions modulate the interlayer hybridization, causing the interference between the two layers' Dirac cones to significantly reconstruct the band structure. This reconstruction gives rise to several notable features: the emergence of secondary Dirac points and minibands, the formation of mini-gaps and Van Hove singularities, suppression of the Fermi velocity at certain twist angles, and the appearance of topologically non-trivial bands.

These effects are especially pronounced near the so-called magic angle ($\sim 1.1°$), where low-energy bands become extremely flat. The resulting enhanced electron correlations can drive phenomena such as superconductivity and Mott-like insulating states [1-3]. Understanding the interplay between interlayer coupling and the Moiré potential is thus central to the physics of TBG.

## 3.2 Magic Angle Phenomenon

The magic angle in TBG refers to specific twist angles at which Fermi velocity $v_F$ near the Dirac point becomes nearly zero due to strong interlayer hybridization, leading to the formation of extremely flat bands and a sharp increase in the density of states [3]. This flat-band condition suppresses kinetic energy and enhances electron–electron interactions, giving rise to correlated phases such as unconventional superconductivity, Mott-like insulating states, and magnetic or nematic ordering [3].

This phenomenon was first predicted by Bistritzer and MacDonald [3], who developed a continuum model showing that interlayer tunneling and momentum mismatch between Dirac cones lead to Fermi velocity renormalization. At the first magic angle (~1.1°), destructive interference in wavefunction overlap flattens the bands, promoting strongly correlated behavior. Experimental confirmation came from Cao *et al.* [1, 2], who observed correlated insulating states and superconductivity in magic angle TBG devices. These breakthroughs established "twistronics" as a new platform for tuning quantum phases through geometric alignment.

Further theoretical and computational studies, including tight-binding and *ab initio* methods [25, 26], have confirmed band flattening near the magic angle and highlighted the essential role of lattice relaxation. Inspired by these findings, other moiré materials such as twisted transition metal dichalcogenides and trilayer graphene have also been explored. Today, twistronics provides a powerful tool for simulating and controlling strongly correlated quantum phenomena driven by geometry and topology [4, 25, 27].

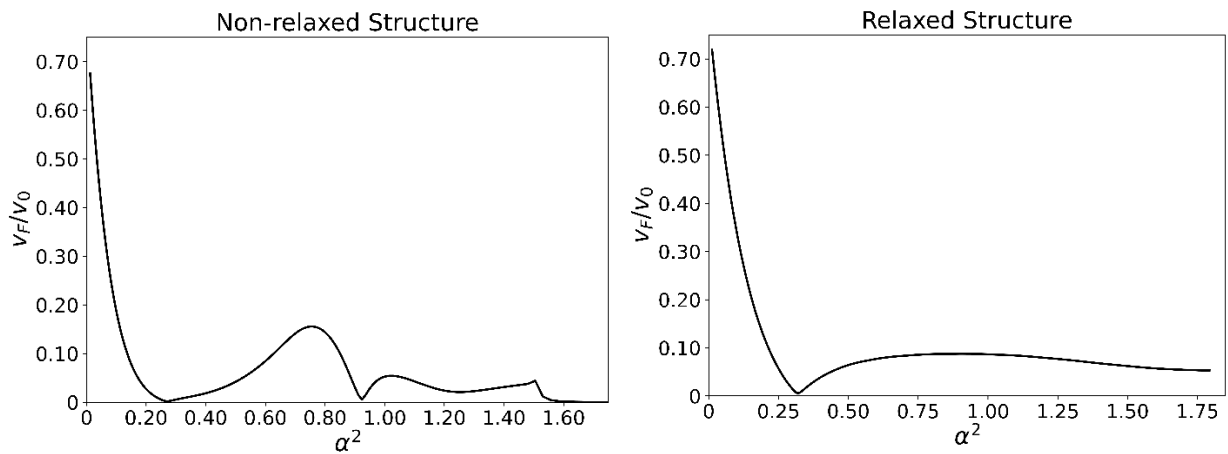

**Figure 3.2.1:** Normalized Fermi velocity ($v_F/v_0$) is plotted against twist parameter $\alpha^2$ ($\propto 1/\theta^2$), normalized to monolayer graphene velocity $v_0$. The relaxation of structure suppresses the presence of second magic angle.

## 3.3 Flat Bands and Superconductivity in Twisted Bilayer Graphene

In twisted bilayer graphene (TBG), flat bands emerge near the Fermi level when the twist angle approaches the magic angle $\theta \sim 1.1°$, leading to a dramatic enhancement of electron–electron interactions [3, 27]. Flat bands imply that electrons have extremely low group velocity and suppressed kinetic energy, which amplifies the effects of Coulomb interactions. This phenomenon arises from quantum interference and interlayer hybridization in the moiré superlattice, producing narrow mini-bands often just a few meV wide near the Fermi energy [2, 5]. This leads to, DoS $N(E_F)$ becomes sharply peaked, and interaction-driven phases dominate. These include correlated insulating behavior at specific fractional band fillings, unconventional superconductivity without phonon mediation, and electronic phases analogous to those in heavy fermion and high $T_c$ cuprate systems [2, 3]. Importantly, such correlated states in TBG are highly tunable via twist angle, doping, strain, or pressure [7, 28].

A major breakthrough came in 2018 when Cao et al. [1, 2] observed superconductivity in magic angle TBG. The proximity of superconductivity to correlated insulators suggests an unconventional pairing mechanism, likely beyond BCS theory. The flat bands enhance $N(E_F)$, fulfilling a key condition for superconductivity even with weak attractive interactions. Theoretical work using ab initio and continuum $\vec{k} \cdot \vec{p}$ models has shown that the emergence of superconductivity is influenced by band topology, interlayer coupling, and lattice relaxation [5, 8]. These findings position TBG as a highly tunable platform for studying strongly correlated and topological superconductivity, offering new insights into quantum phases driven by geometry and electronic interactions.

## 4. Results

The computational results were obtained for twisted bilayer graphene (TBG), focusing on electrical band structure, DoS, and Fermi velocity for various twist angles to explore signatures of twistronics-induced superconductivity, with a focus on appearance of flat bands, enhanced DoS, and suppression of Fermi velocity near the so-called magic angle.

### 4.1 Band Structure Analysis

The calculated band structures for twist angles between 0.80° and 1.35° reveal a progressive evolution of the low-energy moiré bands near the Fermi level $E_F$. For θ = 0.80°–0.93°, the conduction and valence bands remain relatively dispersive, showing no distinct flat-band features. At θ = 0.97°, partial flattening emerges, indicating entry into the pre–magic-angle regime. A pronounced flattening is observed at θ = 0.98°, where the low-energy bands become nearly dispersion less, particularly near the $\Gamma_m$ point along the high-symmetry path. The bands are also relatively isolated from higher-energy states, consistent with enhanced correlation effects. For θ = 0.99° and 1.00°, the bands remain very flat, though a small increase in curvature relative to 0.98° is visible. Beyond this range, at θ = 1.01°–1.02°, dispersion gradually increases, and by θ = 1.10° and 1.35°, the flat-band character is lost, returning to standard bilayer graphene like dispersive behavior.

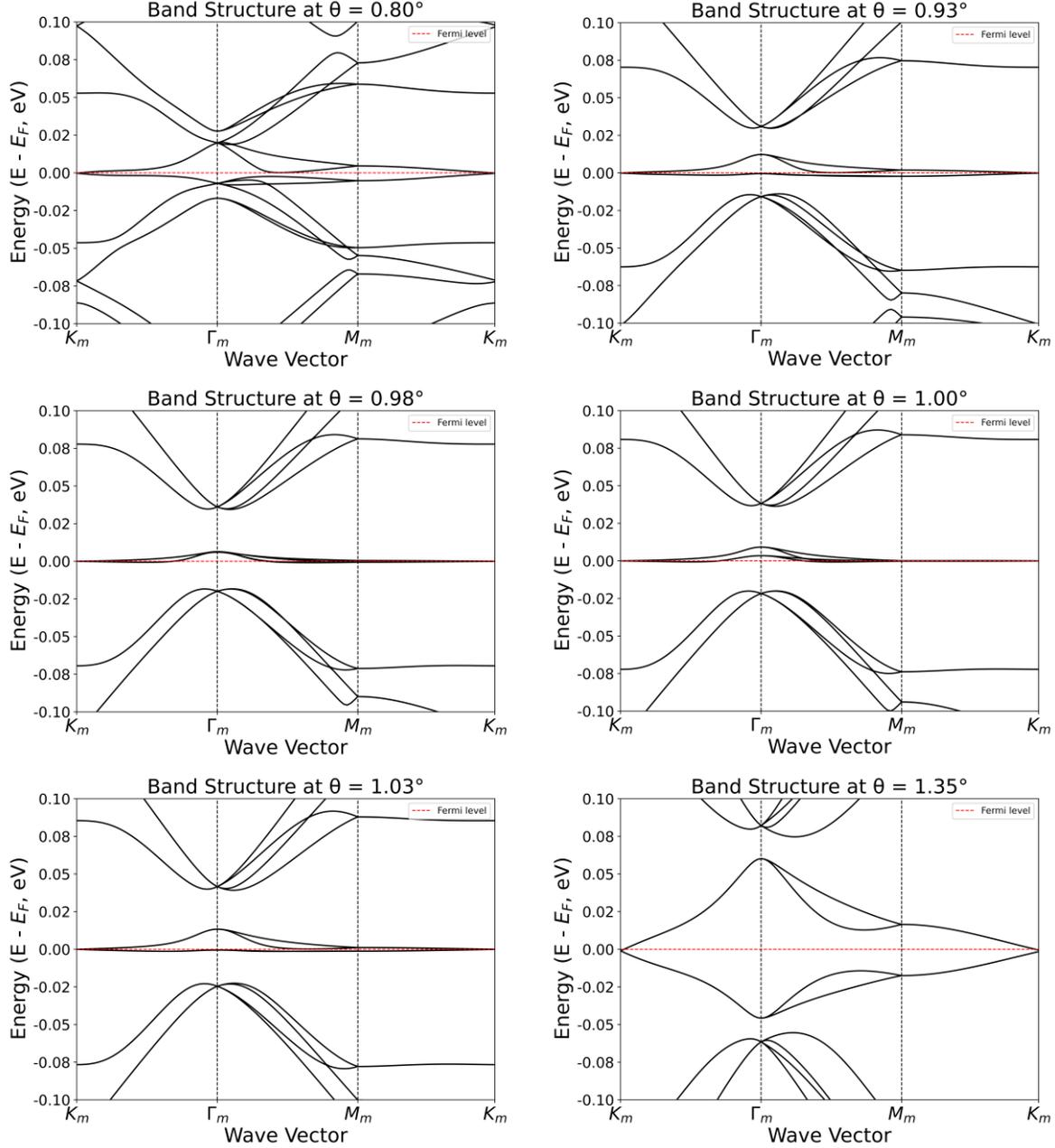

**Figure 4.1.1:** The continuum $\vec{k} \cdot \vec{p}$ model was used to compute the band structure for twist angles: $\theta = 0.80°$ to $1.35°$. The strongest band flattening, which is a sign of greatest correlation effects, occurs at $\theta = 0.98°$.

## 4.2 Density of States Analysis

The DoS plots exhibit a sharp peak at $\theta = 1.00°$, indicating a high density of electronic states at the Fermi level, that is consistent with appearance of flat bands. Importantly, elevated DoS values are also observed at $\theta = 0.98°$ and $1.03°$, reflecting the presence of nearly-flat bands around these twist angles. The high DoS enhances the likelihood of electron pairing interactions, which is a fundamental ingredient for superconductivity. Away from the 0.98°–1.03° range, the DoS at $E_F$ decreases sharply, consistent with the disappearance of flat bands. This trend supports the hypothesis that a narrow range of twist angles around 1.00° creates a conducive environment for correlated phenomena.

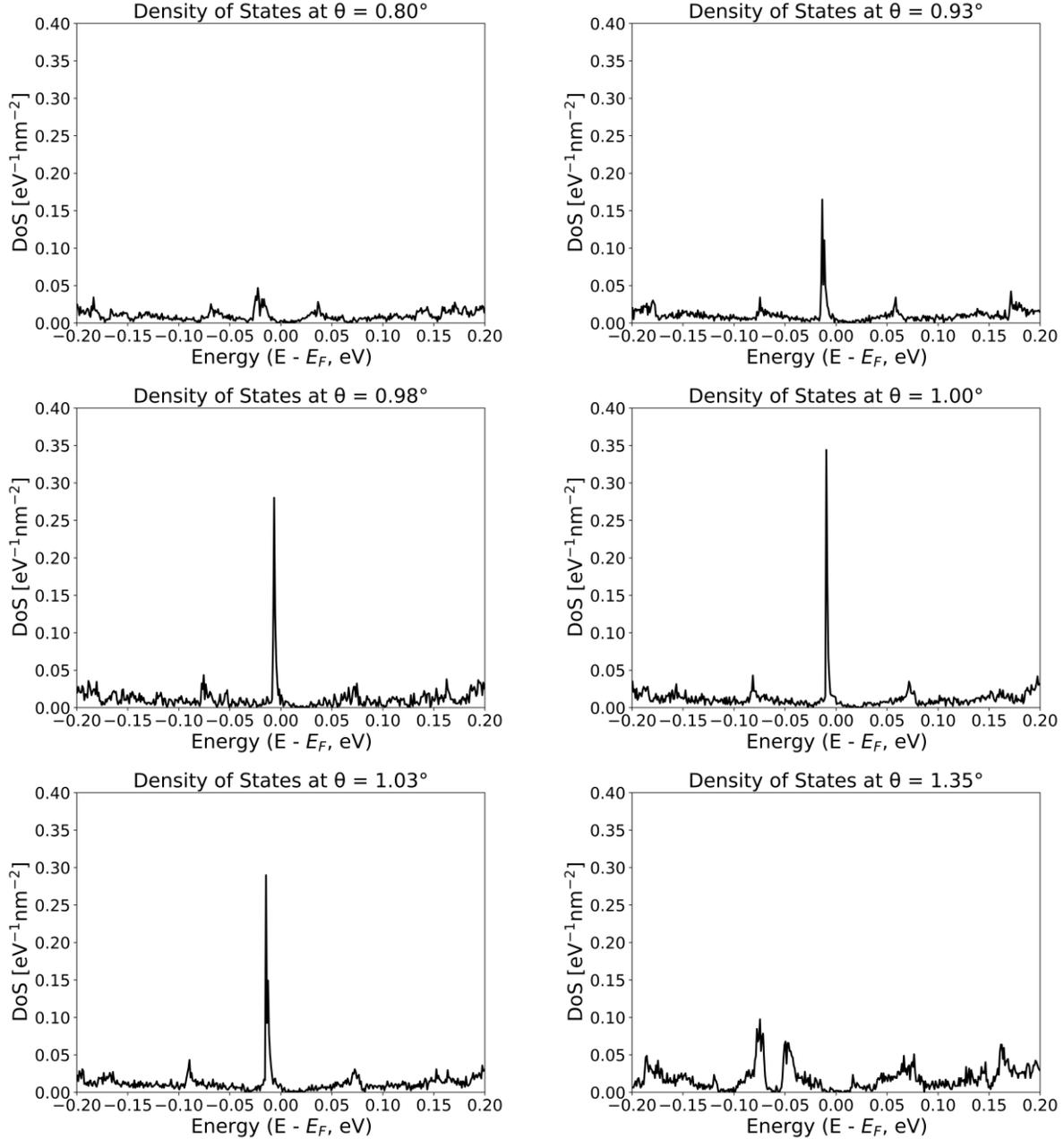

**Figure 4.2.1:** Computed DoS for twist angle from 0.80° to 1.35°. The plot for $\theta = 1.00°$ clearly shows the highest and sharpest peak, while at 0.98° and 1.03° is also highly enhanced.

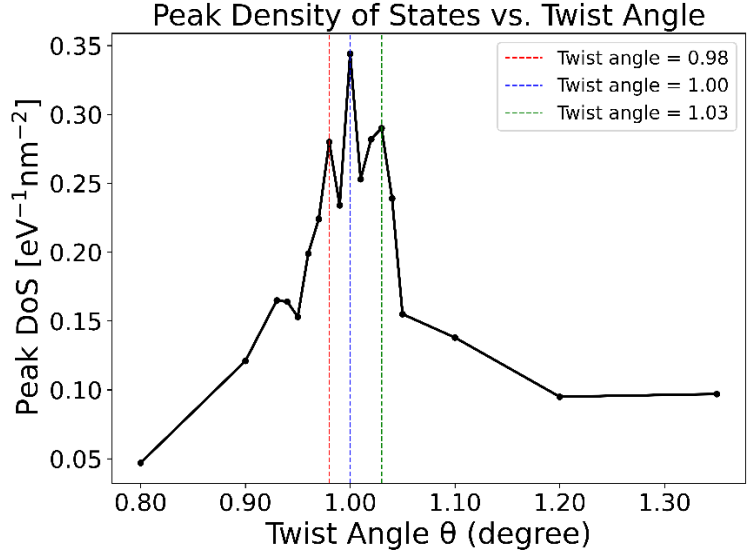

**Figure 4.2.2:** Histogram plot of DoS peak for twist angle between 0.80° to 1.35° which shows that DoS highest peak is at 1.00° and also enhanced at 0.98° and 1.03°.

## 4.3 Fermi Velocity Analysis

The normalized Fermi velocity ($v_F/v_0$), provides further evidence of emergent flat band behavior. The minimum Fermi velocity is observed around θ = 0.98°, 0.99°, and 1.00°, with values approaching zero, particularly in the relaxed structure. Outside this angular range, $E_F$ recovers quickly, consistent with the loss of flatness in the band structure. This dramatic suppression of carrier velocity correlates directly with band flattening, indicating strongly localized electronic states. In the language of BCS superconductivity, reduced Fermi velocity enhances the density of states and facilitates stronger pairing interactions, aligning with the observed DoS peak and flat bands.

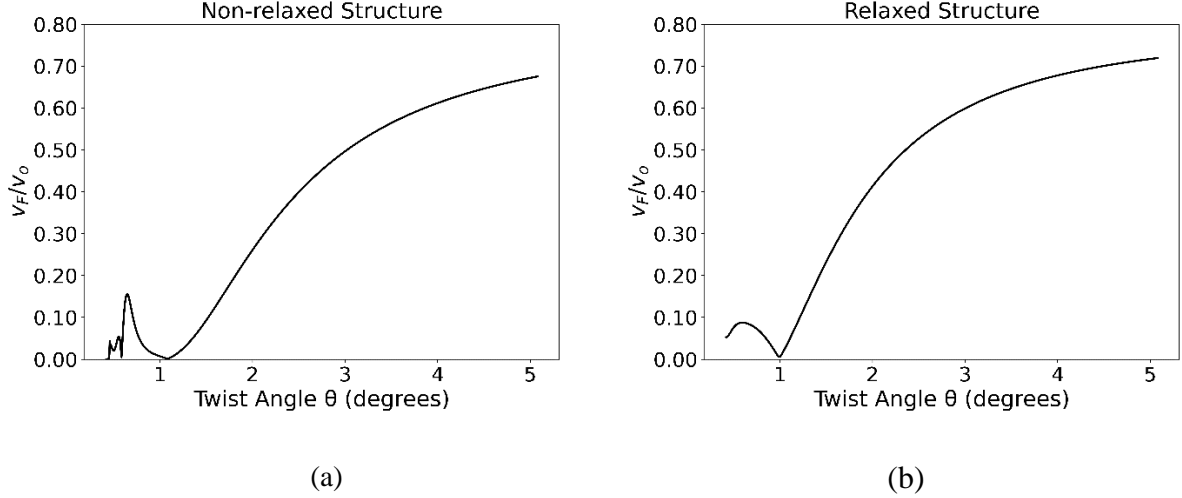

**Figure 4.3.1:** Analysis of the relationship between Fermi velocity $v_F$ and twist angle $\theta$ in TBG for relaxed and non-relaxed configurations: (a) Without lattice relaxation, the Fermi velocity remains relatively high across twist angles. The drop near the magic angle is much less pronounced, emphasizing the crucial part of structural relaxation in development of flat bands. (b) The Fermi velocity $v_F$ significantly drops near the magic angle due to band flattening induced by atomic relaxation. This suppression of $v_F$ is a hallmark of enhanced electronic correlations and flat band formation.

## 5. Discussion

The combined analysis of band structure, DoS, and Fermi velocity identifies a narrow magic-angle window centered around $\theta \approx 0.98°$–$1.00°$. Earlier continuum-model studies by Carr et al. [5] and Koshino et al. [14] broadly reported the first magic angle at ~1.1°, while Cantele et al. [13] emphasized that relaxation is critical for capturing the correct flat-band behavior. Our high-resolution angular sampling refines this picture by revealing that the magic-angle regime is not a sharp point but a narrow interval, as the flattest dispersion occurs at $\theta = 0.98°$, while the highest DoS is found at $\theta = 1.00°$, with elevated values at 0.98° and 1.03°. The minimum Fermi velocities across 0.98°–1.00° indicate a strong suppression of kinetic energy in this range, illustrating that different flat-band signatures may not coincide exactly. The slight mismatch between the flattest bands (0.98°) and the maximum DoS (1.00°) can be explained by the fact that the DoS integrates contributions over the entire Brillouin zone, not just the flattest local points.

This nuanced finding contextualizes experimental observations: correlated insulating and superconducting phases have been shown to appear only within very tight angular tolerances [1,2,10]. Our comparative relaxation analysis further highlights that the non-relaxed structures retain higher Fermi velocities and less pronounced band flattening, the inclusion of relaxation strongly suppresses kinetic energy, sharpening the flat-band condition. These results reinforce the consensus from earlier studies while adding quantitative resolution and unifying multiple observables into a consistent diagnostic framework.

It is important to clarify that while enhanced DoS and suppressed Fermi velocity provide favorable conditions for correlated states, the present continuum model is a single-particle theory and does not directly predict superconductivity. A full microscopic understanding requires incorporating

electron–electron interactions, electron–phonon coupling, or many-body effects, as explored in Hubbard-model and ab initio approaches [7,14,25,27]. Our findings therefore provide necessary groundwork by identifying the angular and structural conditions where such interaction-driven phenomena are most likely to occur.

This work refines and consolidates existing continuum approaches by resolving the magic angle into a narrow 0.98°–1.00° window with high-resolution angular sampling with unifying multiple observables into a consistent flat-band framework. It uses an accessible Python–C++ computational workflow that makes continuum-model studies more reproducible for the broader researchers.

## 6. Conclusion

This study refines the continuum-model description of twisted bilayer graphene by identifying a narrow magic-angle window (0.98°–1.00°) where flat bands, enhanced DoS, and suppressed Fermi velocity coincide. While previous works broadly reported ~1.1° as the first magic angle, our high-resolution angular sweep shows that relaxation sharpens this regime into a much narrower interval, consistent with the extreme alignment sensitivity observed experimentally. The slight mismatch between the angles of maximum DoS (1.00°) and strongest band flattening (0.98°) further illustrates the subtle interplay of different flat-band signatures.

By systematically comparing relaxed and non-relaxed structures, we confirm that atomic relaxation is indispensable for producing the flat-band condition, in agreement with earlier theoretical studies but with added quantitative detail. Our integrated approach consolidates multiple observables into a unified computational framework and offers a replicable workflow for further investigations.

We emphasize that with a critical regime, TBG enters a highly correlated state, an essential electronic prerequisite for correlated phases such as superconductivity, they do not constitute a direct prediction of superconductivity itself rather supports the broader framework of twistronics as a tool for superconductivity in twisted bilayer graphene. Future work incorporating many-body interactions and electron–phonon coupling will be essential for quantitatively connecting flat-band conditions to superconductivity.